\documentclass[twocolumn,fleqn,natbib]{svjour2}
\smartqed 
\usepackage{graphicx,url}
\usepackage{amsmath,amsfonts,amssymb,amsopn,amscd}
\usepackage{mathptmx}
\usepackage[centerlast]{subfigure}
\usepackage{color}
\usepackage{colortbl}
\usepackage{epsfig}

\newcommand{\be}{\begin{equation}}
\newcommand{\ee}{\end{equation}}
\newcommand{\bea}{\begin{eqnarray}}
\newcommand{\eea}{\end{eqnarray}}

\def\url#1{\textcolor{blue}{\underline{#1}}}

\begin{document}


\title{Does brain activity stem from high-dimensional chaotic
dynamics~? Evidence from the human electroencephalogram, cat cerebral
cortex and artificial neuronal networks}

\author{Sami El~Boustani and Alain Destexhe$^*$}

\institute{Unit\'e de Neurosciences Int\'egratives et
Computationnelles (UNIC), \\ CNRS, Gif-sur-Yvette, France \\ $*$:
Corresponding author at the following address: \\ UNIC, Bat 33, CNRS,
1 Avenue de la Terrasse, 91198 Gif sur Yvette, France. \\
\email{Destexhe@unic.cnrs-gif.fr}}

\date{Received: date / Revised: date}

\maketitle


\begin{abstract}

Nonlinear time series analyses have suggested that the human
electroencephalogram (EEG) may share statistical and dynamical
properties with chaotic systems.  During slow-wave sleep or
pathological states like epilepsy, correlation dimension measurements
display low values, while in awake and attentive subjects, there is
not such low dimensionality, and the EEG is more similar to a
stochastic variable.  We briefly review these results and contrast
them with recordings in cat cerebral cortex, as well as with
theoretical models.  In awake or sleeping cats, recordings with
microelectrodes inserted in cortex show that global variables such as
local field potentials (local EEG) are similar to the human EEG. 
However, in both cases, neuronal discharges are highly irregular and
exponentially distributed, similar to Poisson stochastic processes. 
To attempt reconcile these results, we investigate models of
randomly-connected networks of integrate-and-fire neurons, and also
contrast global (averaged) variables, with neuronal activity.  The
network displays different states, such as ``synchronous regular''
(SR) or ``asynchronous irregular'' (AI) states.  In SR states, the
global variables display coherent behavior with low dimensionality,
while in AI states, the global activity is high-dim\-ensionally chaotic
with exponentially distributed neuronal discharges, similar to awake
cats.  Scale-dependent Lyapunov exponents and $\epsilon$-entropies show that the
seemingly stochastic nature at small scales (neurons) can coexist
with more coherent behavior at larger scales (averages).  Thus, we
suggest that brain activity obeys similar scheme, with seemingly
stochastic dynamics at small scales (neurons), while large scales
(EEG) display more coherent behavior or high-dimensional chaos.

\end{abstract}

\section{Introduction}

A number of methods from nonlinear dynamical systems have been
applied to the different states of the human EEG.  Early studies have
calculated correlation dimensions from EEG and reported evidence for
low-dimensional chaos for slow-wave sleep [Babloyantz et al., 1985;
Mayer-Kress et al., 1988], as well as for pathological states such as
epilepsy [Babloyantz \& Destexhe, 1986; Frank et al., 1990] or the
terminal state of Creutzfeldt-Jakob disease [Destexhe et al., 1988]. 
These findings have been confirmed by numerous studies [reviewed in
Destexhe, 1992; Elbert et al., 1994; Korn \& Faure, 2003]. 
Interestingly, in these studies, the EEG dynamics during wakefulness
or REM sleep did not show evidence for low-dimensional dynamics. 
These results were also corroborated with other measurements such as
Lyapunov exponents, entropies, power spectral densities,
autocorrelations and symbolic dynamics [Destexhe, 1992].

The existence of low-dimensional chaotic dynamics in systems with
such large degrees of freedom is surprising.  Chaotic dynamics in
extended coupled systems have been studied extensively for the last
decades and is still a matter of intense investigation.  In
particular, for network of simplified neurons, as the spin-glass
models, mean-field theories have proven the existence of stable chaotic
attractors [Sompolinsky et al., 1988; van Vreeswijk \& Sompolinsky,
1996; van Vreeswijk \& Sompolinsky, 1998]. These systems have been
shown to exhibit chaotic persistence regarding parameter changes
[Albers et al., 2006], a property that makes chaotic dynamics more
common than exceptional [Sprott, 2008]. From a computational point of
view, chaotic behaviour or nearly chaotic regime (edge of chaos) can
be optimal for information processing [Bertschinger \& Natschl\"ager,
2004; Legenstein \& Maass, 2007]. In network models of spiking
neurons, chaotic regimes have also been studied [Cessac, 2008; 
Cessac \& Vi\'eville, 2008] and can be present
only as transients during which the system is numerically
indistinguishable from a usual chaotic attractor.  However the
lifetime of these transient periods is known to increase
exponentially with network size [Crutchfield \& Kaneko, 1988; T\'el
\& Lai, 2008] making them more relevant in practice than the real
stable attractor for large enough networks.  

In this paper, we intend to study a specific spiking neuron model
which displays these properties and yield biophysical behaviour
relevant to understand EEG data. Classical nonlinear tools as the
correlation dimension or the Lyapunov exponents have given some
insight on macroscopic quantities as the EEG or the overall activity
of numerical models but are often criticized because of their
limitation for low-dimensional dynamics [Kantz \& Schreiber, 2004]. 
In order to distinguish between microscopic dynamics and collective
behaviour, we borrow recent tools developed in the context of
high-dimensional systems and which offer analysis at different
scales.  In particular, finite-size Lyapunov exponent [Aurell et al.,
1997] as well as $\epsilon-$entropy [Gaspard \& Wang, 1993] provide a
scale-dependent description of spiking neuron networks and could
detect whether or not low-dimensional dynamics prevail at a
macroscopic scales. These tools will be applied to EEG data as well
as numerical models.

\section{Methods}

In this section we briefly describe the analytical and numerical
tools which will be used to probe the nonlinear nature of the EEG
recordings or the numerical simulations. Most of these analysis are
extensively used in the literature. Furthermore, we used the TISEAN
toolbox [Hegger et al., 1999] to perform most of our nonlinear
analysis. We also give information about the numerical model.

\subsection{Phase space reconstruction}

We used the method of time-delayed vectors of the time series, which
yields reconstructed attractors topologically equivalent to the
original attractor of the system [Takens, 1981; Sauer et al., 1991]. 
We chose a fixed delay parameter determined by the first minimum of
the mutual information [Fraser \& Swinney, 1986].  The embedding
dimension was chosen such that any self-similar asymptotic behavior
saturates beyond this dimension, indicating a successful attractor
reconstruction.

\subsection{Correlation dimension}

The correlation dimension was measured from the correlation integral
[Grassberger \& Procaccia, 1983] which is estimated by using 
\begin{equation}
C(\epsilon) = \frac{1}{N^2}\sum_{i,j=1}^{N}\Theta(\epsilon - ||\textbf{x}(i)-\textbf{x}(j)||)
\end{equation}
where $\textbf{x}(i) \in 
\mathrm{I\!R}$
is the m-dimensional delay vector and $\epsilon$ is a threshold
distance which reflects the scale under consideration. Moreover we
used a Theiler window [Theiler, 1990] according to the temporal
correlation of each time series to avoid spurious estimation. If the
correlation integral manifests a power-law behavior over a range of
scale which saturates with increasing embedding dimension, the
correlation dimension is simply the corresponding power-law
coefficient.

\subsection{Recurrence plots}

This tool provides an intuitive visualization of recurrences in
trajectories on the attractor [Eckmann et al., 1987]. The
recurrence matrix is defined by 
\begin{equation}
RP(i,j) = \Theta(\epsilon - ||\textbf{x}(i)-\textbf{x}(j)||)
\end{equation}
where the matrix entry is equal to 1 if the trajectory $\textbf{x}$ 
at $i$ is in the $\epsilon$-neighbor of itself at $j$ and 0 
otherwise. Stereotypic patterns in the recurrence plot can indicate 
the existence of periodic behavior, low-dimensional chaos or 
stochastic-like dynamics. In particular, diagonal lines are typical 
of periodic dynamics whereas clouds of dots are produced by a 
stochastic component. For an extensive review on recurrence plot, 
the reader should refer to [Marwan et al., 2006].

\subsection{Finite-size Lyapunov and $\epsilon$-entropy}

Scale-dependent nonlinear analysis have been greatly studied the last
decade and have come in handy to distinguish deterministic (possibly
chaotic) dynamics from stochastic noise. In this paper, we will
mainly consider two quantities : the Finite-Size Lyapunov Exponent (FSLE) and
the $\epsilon$-entropy. The former has been introduced in the context
of developed turbulence [Aurell et al., 1997] and has proven to be
more suited for a broad range of system which dynamics can exhibit
low-dimensional chaotic behavior only on large-scale [Shibata \&
Kaneko,1998; Cencini et al., 1999;  Gao et al., 2006]. Roughly
speaking, if we want to quantify the sensitivity to initial conditions
on large scales, it is necessary to consider perturbations which are not
infinitesimal. Therefore, the FSLE can be defined as follows
\begin{equation}
\lambda(\delta) = \frac{1}{<T_{r}(\delta)>}\left\langle ln\left( \frac{\Delta_r}{\delta} \right) \right\rangle\label{eq:FSLE}
\end{equation}
where for a perturbation between two initial trajectories $\delta$, 
$T_r$ is the minimal time required for those trajectories to be 
separated by a distance $\Delta_r$ greater than or equal to $\delta r$ where 
$r=2$ is usually taken. The brackets in Eq.~\ref{eq:FSLE} denotes 
averages on the attractor for many realizations. The $\epsilon$-entropy is a generalization 
of the Kolmogorov-Sinai entropy rate [Gaspard \& Wang, 1993] which is 
defined for a finite scale $\epsilon$ and time delay $\tau$ by
\begin{eqnarray}
h(\epsilon,\tau) & = & \lim_{m\rightarrow\infty}h_{m}(\epsilon,\tau)\\
 & = & \frac{1}{\tau}\lim_{m\rightarrow\infty}\frac{1}{m}H_{m}(\epsilon,\tau)
\end{eqnarray}
with 
\begin{equation}
h_{m}(\epsilon,\tau) = \frac{1}{\tau}(H_{m+1}(\epsilon,\tau)-H_{m}(\epsilon,\tau)
\end{equation}
where $H_{m}(\epsilon,\tau)$ is the entropy estimated with box partition of the phase space for box size 
given by $\epsilon$ on the attractor reconstructed with a time delay 
$\tau$ and an embedding dimension $m$. This quantity exhibits a 
plateau on particular scales if a deterministic low-dimensional 
dynamics occurs at these scales. It can thus be used to describe 
large scale dynamics independently of the small scale noisy behavior 
or can be used to distinguish noise from chaos in some cases 
[Cencini, 2000].

\subsection{Avalanche analysis}

To identify the presence of scale invariance, typical of
self-organized critical states [Jensen, 1998], we used an ``avalanche
analysis'' methods [Beggs and Plenz, 2003].  This method consists of
detecting ``avalanches'' as clusters of contiguous events separated
by silences, by binning the system's activity in time windows (1~ms
to 16~ms were used).  Cluster of events were defined from the spike
times among the ensemble of simultaneously recorded neurons.  The
scale invariance was determined from the distribution of avalanche
size, calculated as the total number of events [Beggs and Plenz,
2005].  

\subsection{Network of Integrate-and-Fire neurons}

Networks of "integrate-and-fire" neurons were simulated according to
models and parameters published previously [Vogels \& Abbott, 2005;
El Boustani et al., 2007].  The network consisted of 5,000 neurons,
 which were separated into two populations
of excitatory and inhibitory neurons, forming 80\% and 20\% of the
neurons, respectively.  All neurons were connected randomly using a
connection probability of 2\%.

The membrane equation of cell $i$ was given by:
\be
 C_m \ {dV_i \over dt} \ = \ -g_L (V_i-E_L) \ + \ S_i(t) \ + \ G_i(t) ~ ,
\ee
where $C_m$ = 1~$\mu$F/cm$^2$ is the specific capacitance, $V_i$ is
the membrane potential, $g_L$ = 5$\times$10$^{-5}$~S/cm$^2$ is the
leak conductance density and $E_L$ = -60~mV is the leak reversal
potential.  Together with a cell area of 20,000~$\mu$m$^2$, these
parameters give a resting membrane time constant of 20~ms and an
input resistance at rest of 100~M$\Omega$.  The function $S_i(t)$
represents the spiking mechanism intrinsic to cell $i$ and $G_i(t)$
stands for the total synaptic current of cell $i$.  Note that in this
model, excitatory and inhibitory neurons have the same properties.

In addition to passive membrane properties, IF neurons had a firing
threshold of -50~mV.  Once the Vm reaches threshold, a spike is
emitted and the membrane potential is reset to -60~mV and remains at
that value for a refractory period of 5~ms.  This model was inspired
from a previous publication reporting self-sustained irregular states
[Vogels and Abbott, 2005].

Synaptic interactions were conductance-based, according to the
following equation for neuron $i$:
\be
 G_i(t) \ = \ - \ \sum_{j} g_{ji}(t) (V_i-E_j) ~ ,	\label{gcond}
\ee
where $V_i$ is the membrane potential of neuron $i$, $g_{ji}(t)$ is
the synaptic conductance of the synapse connecting neuron $j$ to 
neuron $i$, and $E_j$ is the reversal potential of that synapse.  
$E_j$ was 0~mV for excitatory synapses, or -80~mV for inhibitory
synapses.  

Synaptic interactions were implemented as follows: when a spike
occurred in neuron $j$, the synaptic conductance $g_{ji}$ was
instantaneously incremented by a quantum value ($q_e$ = 6~nS and
$q_i$ = 67~nS for excitatory and inhibitory synapses, respectively)
and decayed exponentially with a time constant of $\tau_e$ = 5~ms and
$\tau_i$ = 10~ms for excitation and inhibition, respectively.

\section{Evidence for chaotic dynamics in EEG activity}
\label{EEGsec}

\begin{figure}[h] 
\centerline{\psfig{figure=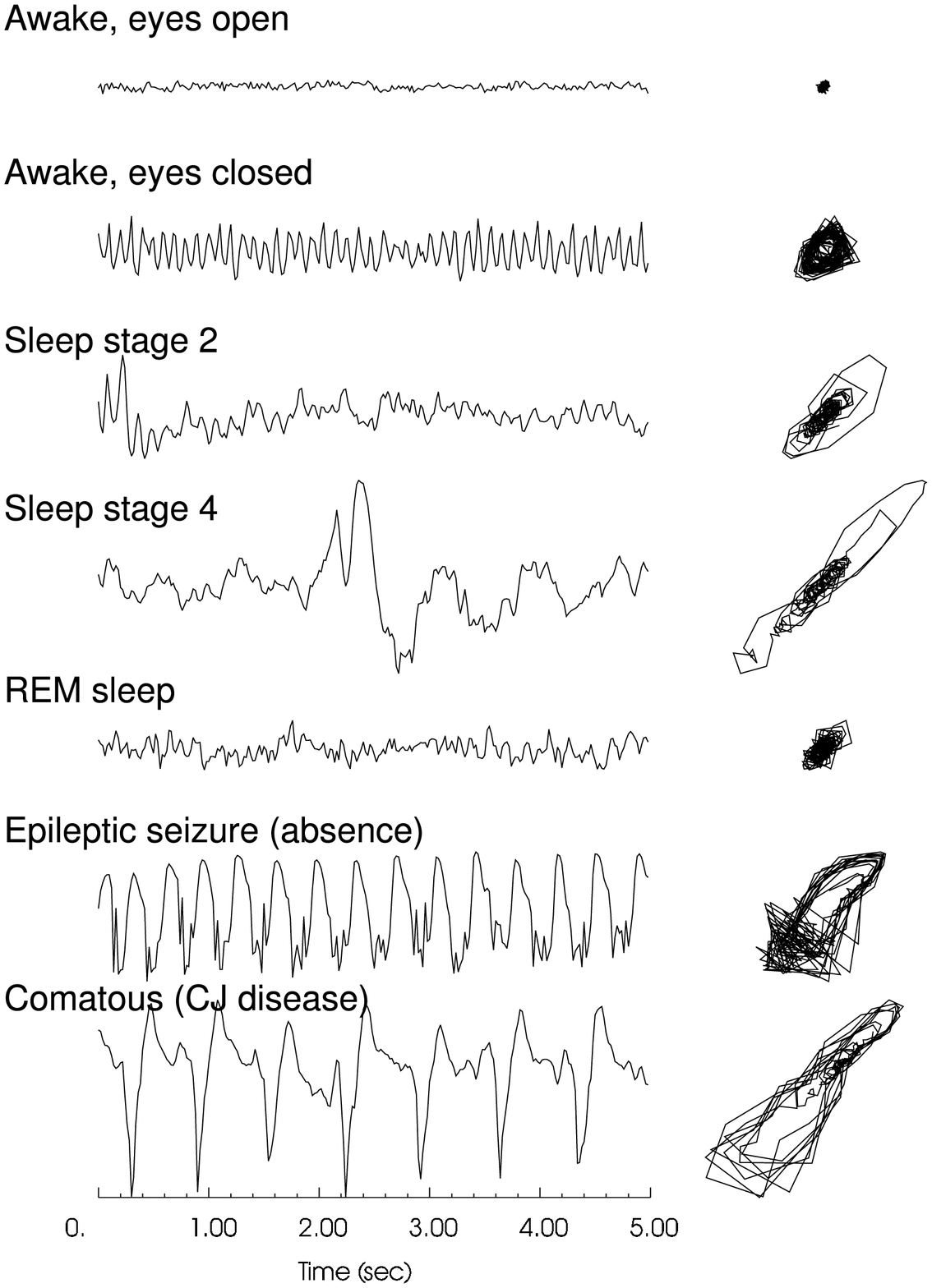,width=0.9\columnwidth}}

\caption{Electroencephalogram signals and phase portraits during
different brain states in humans.} \label{eeg} \small Left: 5
seconds of EEG activity in different brain states (same amplitude
scale).  Right: 2-dim phase portrait of each signal.  Modified from
[Destexhe, 1992].

\end{figure}

Human EEG recordings during different brain states are illustrated
in Fig.~\ref{eeg}, along with a 2-dimensional representation of the
phase portrait obtained from each signal.  During wakefulness (eyes
open) and REM sleep, the dynamics is characterized by low-amplitude
and very irregular EEG activity, while during deep sleep, the EEG
displays slow waves (``delta waves'') of large amplitude. 
Oscillatory dynamics with a frequency around 10~Hz is seen in the
occipital region when the eyes are closed (``alpha rhythm'').
Pathological states, such as epilepsy or comatous states, display
large amplitude oscillations, which are strikingly regular.

\begin{figure}[h] 
\centerline{\psfig{figure=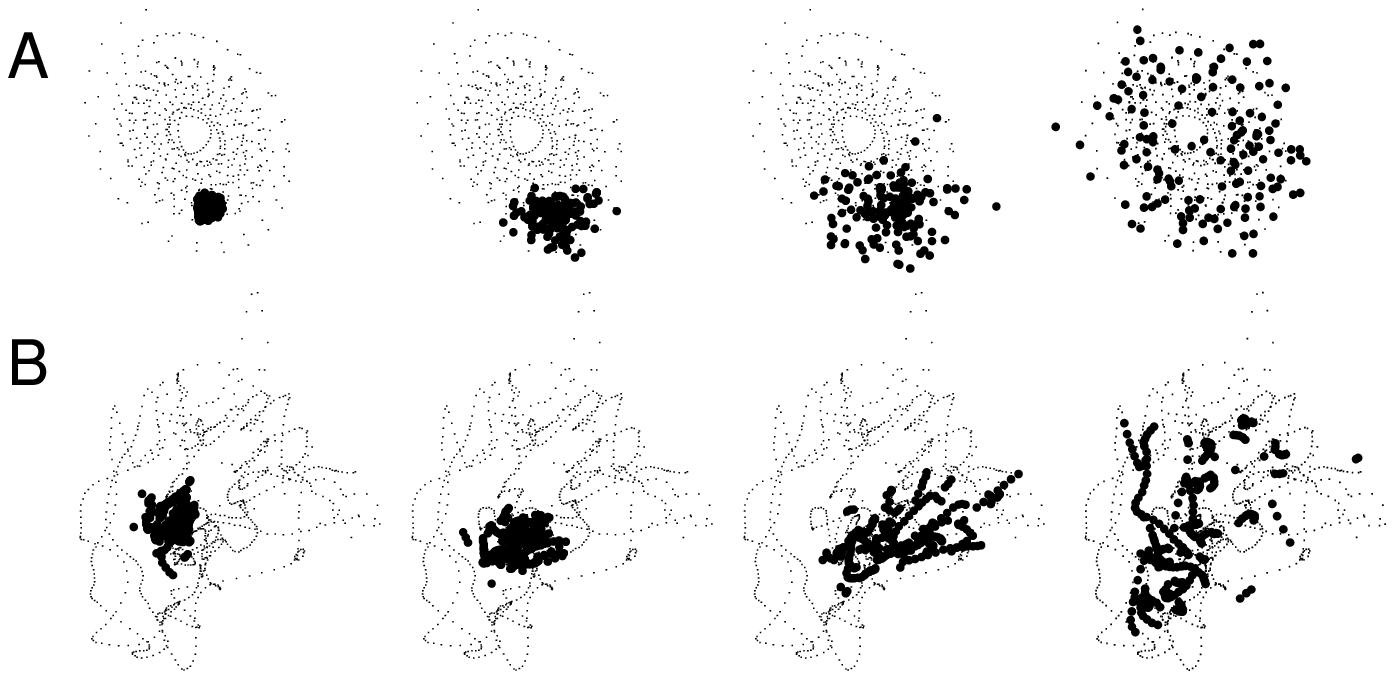,width=0.9\columnwidth}}

\caption{Illustration of the sensitivity to initial conditions in EEG
dynamics.} \label{sensit}  A cluster of neighboring points in phase
space (leftmost panels) is followed in time and is shown on the same
phase portrait after 200~ms, 400~ms and 3 seconds (from left to
right; same data as in Fig.~\ref{eeg}).  A, Awake eyes closed; B.
Sleep stage IV.   Modified from [Destexhe, 1992].

\end{figure}

A first evidence for chaotic dynamics is that EEG dynamics display a
prominent sensitivity to initial conditions.  This sensitivity is
illustrated in Fig.~\ref{sensit} for the alpha rhythm (awake eyes
closed) and slow-wave sleep (stage IV).  A set of close initial
conditions is defined by choosing a neighboring points in phase
space, and this set of points is followed in time.  The divergence of
the trajectories emanating from each initial condition is evident
from the illustration of Fig.~\ref{sensit}, and is actually
exponential.  This exponential divergence betrays the presence of at
least one positive Lyapunov exponent.  A more quantitative
investigation using numerical methods [Wolf et al., 1985] reveals the
presence of positive Lyapunov exponents for all EEG states
investigated [Destexhe, 1992].

\begin{figure}[h] 
\centerline{\psfig{figure=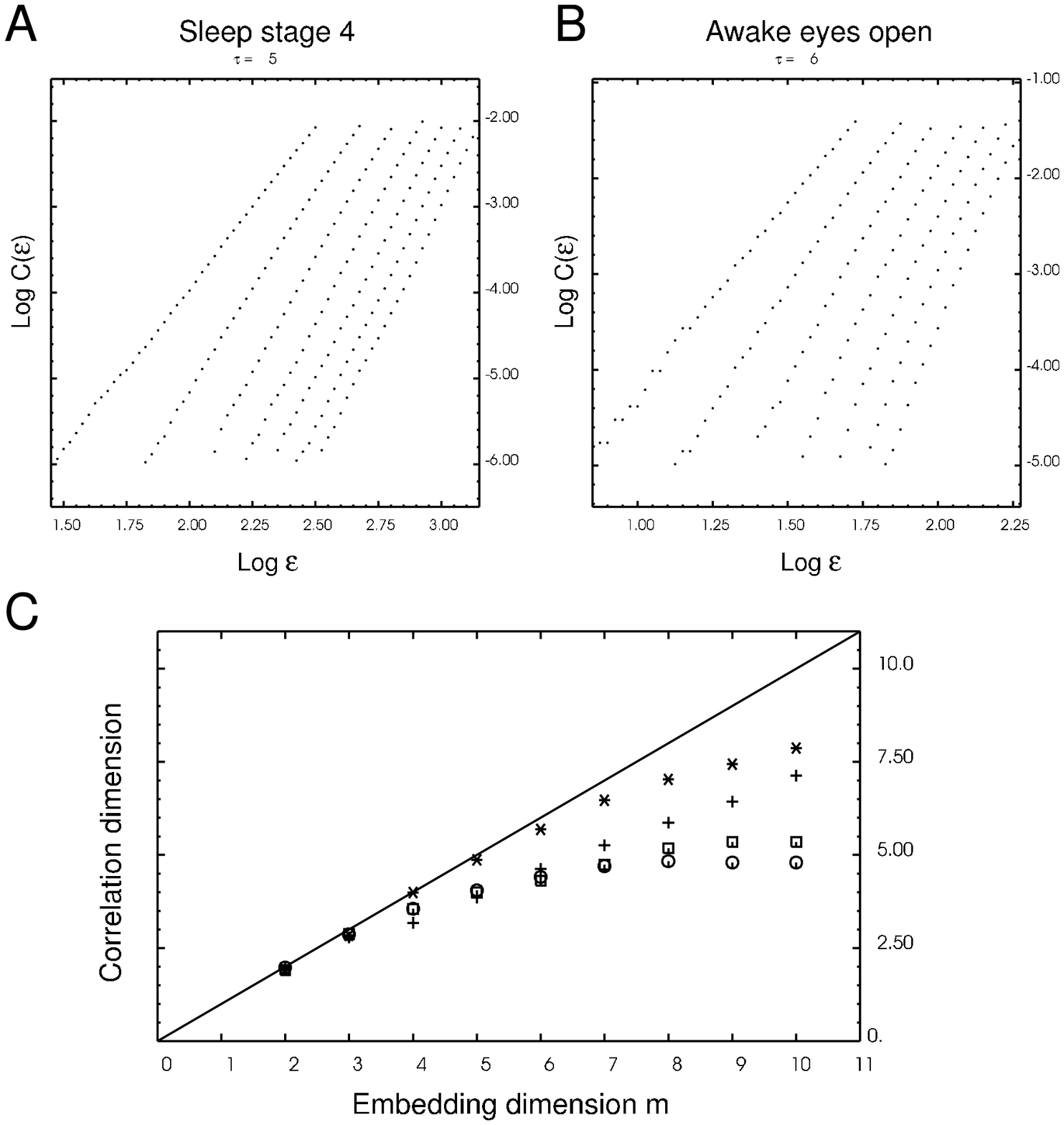,width=0.9\columnwidth}}

\caption{Correlation dimension calculated from different brain states
in humans.} \label{cdim} \small A. Correlation integrals $C(r)$
calculated from sleep stage IV.  B. Correlation integrals calculated
from Awake eyes open.  C.  Correlation dimension as a function of
embedding dimension for different brain states.  Symbols: * = Awake
eyes open, + = REM sleep, squares = sleep stage 2, circles = sleep
stage 4.  Modified from [Destexhe, 1992].

\end{figure}

EEG dynamics also displays other characteristics of chaotic dynamical
systems, such as broad-band power spectra (not shown) and fractal
attractor dimensions.  This latter point was investigated by a number
of laboratories, and is summarized in Fig.~\ref{cdim}.  Correlation
integrals $C(r)$ are calculated from the reconstructed phase
portraits for different embedding dimensions (Fig.~\ref{cdim}A-B). 
The correlation dimension $d$ is obtained by estimating the scaling
of $C(r)$ by using logarithmic representations, in which the slope
directly gives an estimate of $d$.  The calculation of $d$ as a
function of the embedding dimension (Fig.~\ref{cdim}C) saturates to a
constant values for some states (such as deep sleep or pathologies),
but not for others.  This is the case for the EEG during wakefulness,
which dimension $d$ does not saturate, which is a sign of high
dimensionality.

\begin{figure}[h] 
\centerline{\psfig{figure=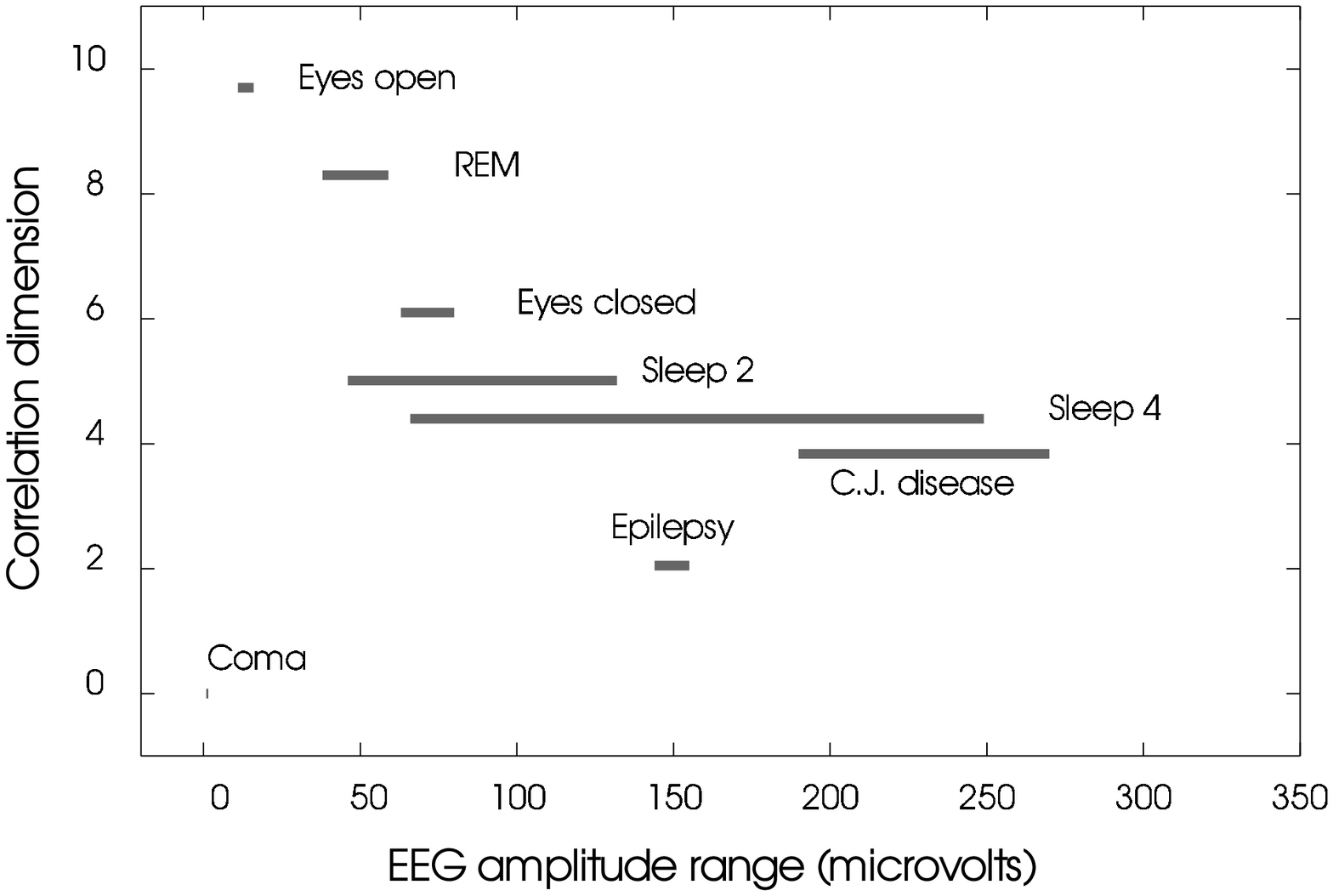,width=0.9\columnwidth}}

\caption{Dimension -- amplitude representation for different brain
states.} \label{dimampl}  The correlation dimension of the EEG is
shown as a function of the amplitude range (maximal amplitude
deflection calculated over 1~second periods.  Modified from
[Destexhe, 1992].

\end{figure}

The correlation dimensions obtained for different brain states are
represented in Fig.~\ref{dimampl} as a function of the mean amplitude
of the EEG signal.  This representation reveals a ``hierarchy'' of
dimensionalities for the EEG.  The aroused states, such as
wakefulness and REM sleep, are characterized by high dimension and no
sign of slow-wave activity.  As the brain drifts towards sleep, the
dimensionality decreases and attains its lowest level during the
deepest phase of sleep, in which the EEG is dominated by
large-amplitude slow waves.  A further decrease is seen during
pathologies, which are also dominated by slow-wave activity.

\section{Evidence for stochastic dynamics in brain activity}

\begin{figure}[h] 
\centerline{\psfig{figure=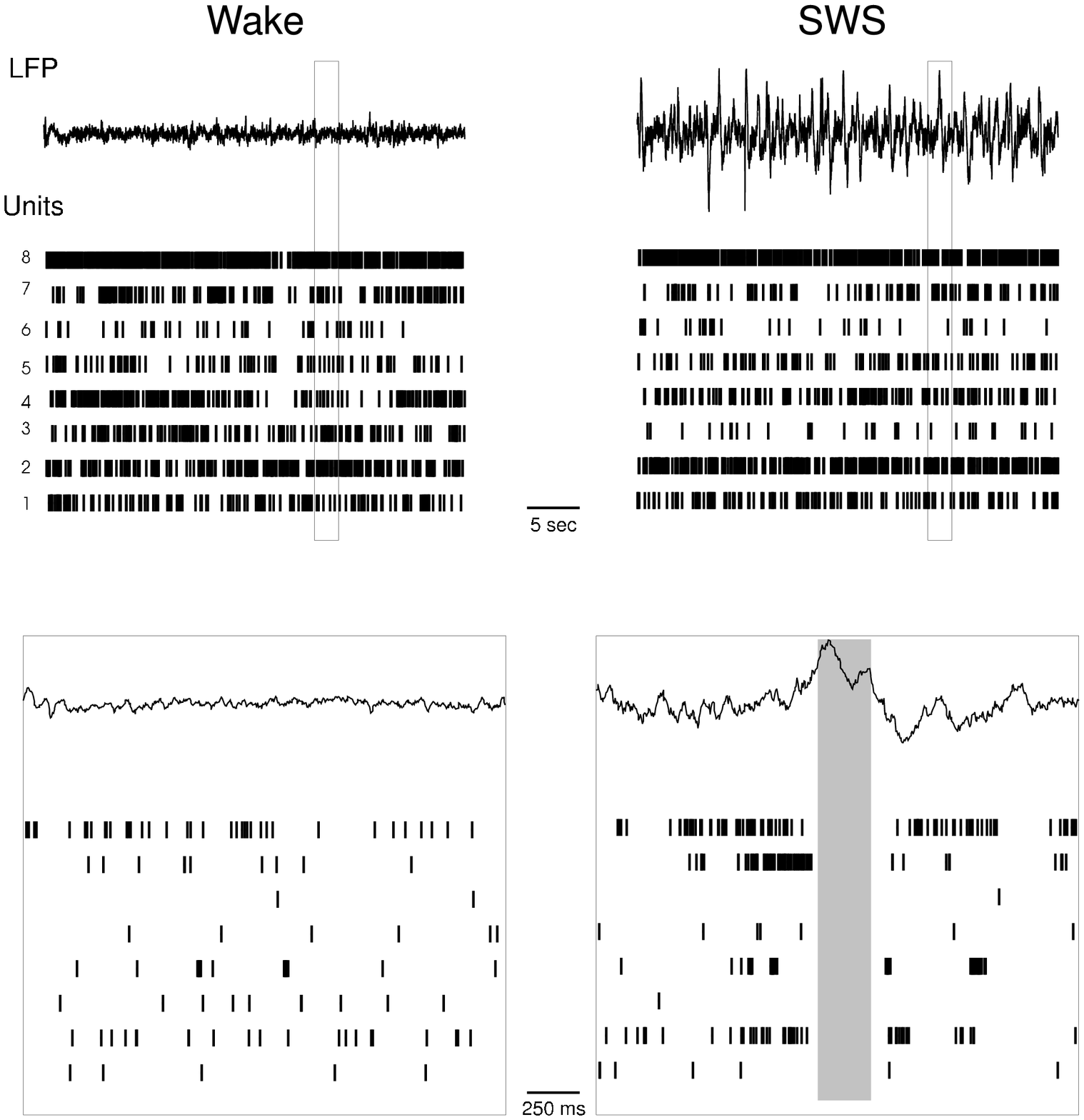,width=1.0\columnwidth}}

\caption{Distributed firing activity and local field potentials in
cat cortex during wake and sleep states.} \label{lfp-spikes}
Recordings were done using eight bipolar tungsten electrodes in cat
parietal cortex (data from [Destexhe et al., 1999]).  The irregular
firing activity of 8 multi-units is shown at the same time as the LFP
recorded in electrode 1.  During wakefulness, the activity is
sustained and irregular (see magnification below).  During slow-wave
sleep (SWS), the activity is similar as wakefulness, except that
``pauses'' of firing occur in all cells, and in relation to the slow
waves (one example is shown in gray in the bottom graph).  The boxes
in the top graphs are shown in bottom graphs at 20 times higher
temporal resolution.

\end{figure}

The above results are consistent with the idea that awake brain
activity may be associated with high-dimensional dynamics, perhaps
analogous to a stochastic system.  To further investigate this
aspect, we have examined data from animal experiments in which both
microscopic (cells) and macroscopic (EEG) activities can be recorded.
The correspondence between these variables is shown in
Fig.~\ref{lfp-spikes} for cat cerebral cortex during wakefulness and
slow-wave sleep.

Those electrical measurements were made using tungsten
microelectrodes directly inserted in cortical gray matter [Destexhe
et al., 1999].  This recording system enables the extraction of two
signals: a global signal, similar to the EEG, which is called ``local
field potential'' (LFP) and reflects the averaged electrical activity
of a large population of neurons.  In addition, single neurons can be
distinguished and can be extracted.  These signals are shown and
compared in Fig.~\ref{lfp-spikes}.  During wakefulness, the LFP shows
low-amplitude irregular activity, while neuronal discharges seem
random. Slow-wave sleep is characterized by dominant delta waves in
the LFP, as in human sleep stage IV.  The occurrence of slow waves is
correlated with a concerted pause in the firing of the neurons
[Destexhe et al., 1999].  

\begin{figure}[h] 
\centerline{\psfig{figure=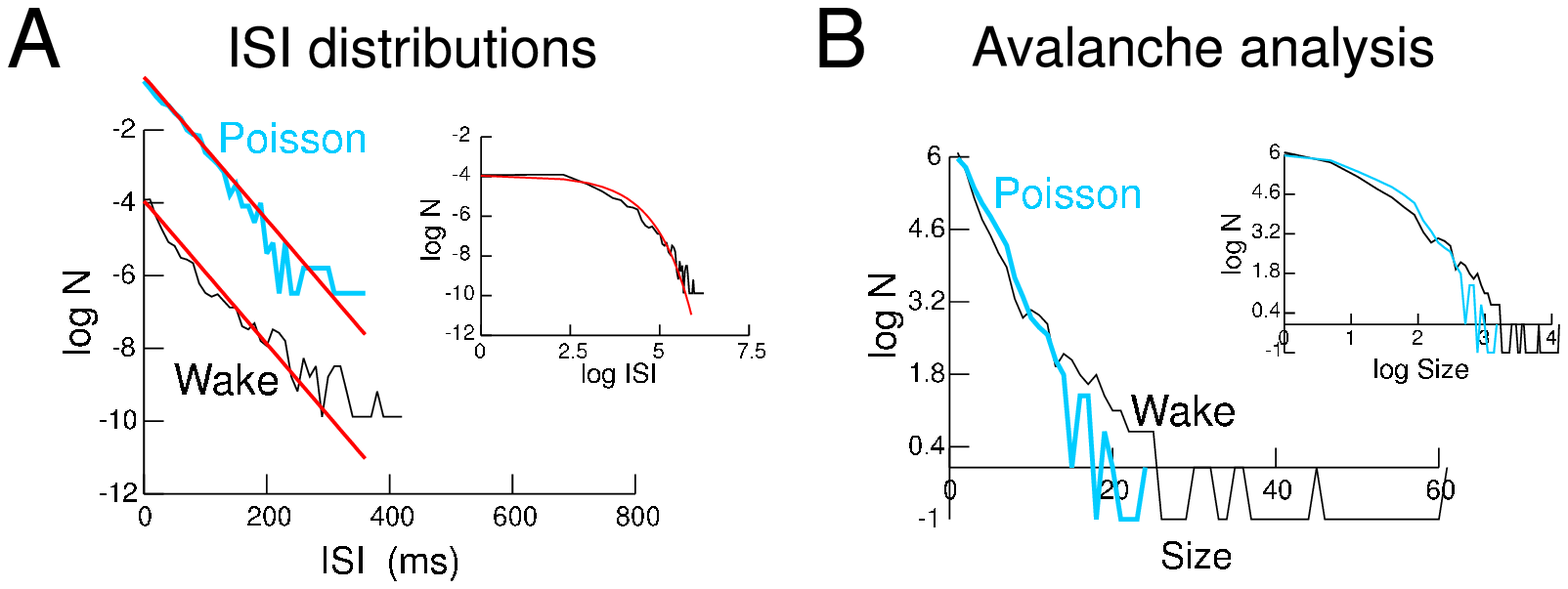,width=1\columnwidth}}

\caption{Analysis of neuronal activity in awake cat cerebral cortex.}
\label{anal} A. Interspike interval (ISI) distributions computed from
extracellularly recorded neurons in wakefulness (natural logarithms).
A Poisson process with the same average rate is shown for comparison
(blue; curve displaced upwards for clarity).  The inset shows a
log-log representation. The red curves indicate the theoretical value
for Poisson processes.  B.  Avalanche analysis of extracellular
recordings in the awake cat (natural logarithms).  The same analysis
was performed on surrogate data (blue; Poisson processes).  The inset
shows the same data in log-log representation.  Modified from Bedard
et al., 2006.

\end{figure}

Analyzing the spike discharge of single neurons revealed that the
interspike-intervals (ISI) are exponentially distributed, in a manner
indistinguishable from a Poisson stochastic process
(Fig.~\ref{anal}A).  Performing an avalanche analysis revealed that
the distribution of avalanche size from the neuronal discharges was
also exponential (Fig.~\ref{anal}B).  The same scaling could be
explained by uncorrelated Poisson processes, as if the neurons
discharged randomly and independently.  This analysis was reproduced
from a previous study (Bedard et al., 2006). 

Thus, these data and analysis show that the dynamics of neuronal
activity in the awake cerebral cortex is similar to stochastic
processes.  The statistics of the ISI distributions, as well as the
collective dynamics (``avalanches'') cannot be distinguished from a
Poisson stochastic processes.

\section{Models of irregular dynamics in neuronal networks}

\begin{figure}[h] 
\centerline{\psfig{figure=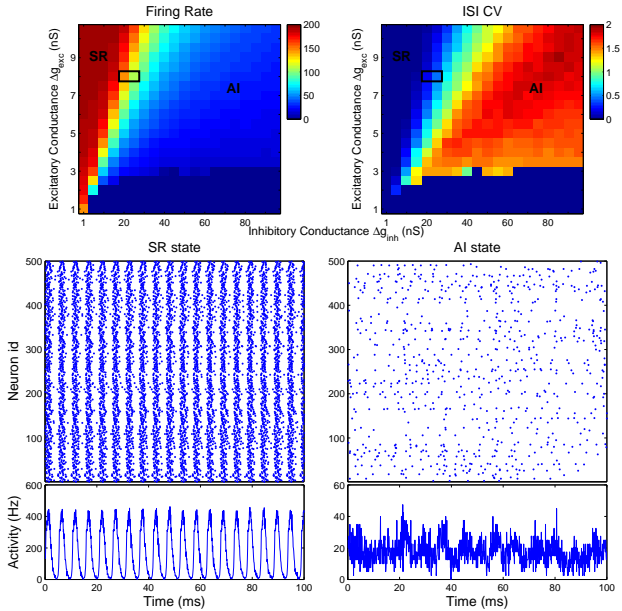,width=1.0\columnwidth}}

\caption{Different network states in randomly-connected networks of
spiking neurons.} \label{network} Top: state diagrams representing
the average firing rate (left) and the firing irregularity (ISI CV,
right) of a randomly-connected network of 5,000 integrate-and-fire
neurons with a sparse connectivity of 2\%. The phase diagram is drawn
according to the excitatory and inhibitory quantal conductance. For
increasing inhibition the dynamics undergo a transition from
synchronous firing among the neuron and regular firing to a state
where the synchrony has substantially decreased and neurons fire
irregularly. {The black rectangle indicates a transition from
the SR to the AI regime. We selected this region for the rest of the
study (see next figures).} Bottom: the two different states,
``Synchronous Regular'' (SR, left) and ``Asynchronous Irregular''
(AI, right).  The corresponding regions in the state diagram are
indicated on top. For both panels, the upper part show a raster plot
of a 500 neurons taken randomly from the network. Each dot is a spike
emitted by the corresponding neuron across time. In the lower part,
the mean firing rate computed among the whole network with a time bin
of 0.1 ms. The asynchronous activity is reflected through the less
fluctuating mean activity.

\end{figure}

We next consider if this type of dynamics can be found in theoretical
models of neuronal networks.  We consider ran\-domly-connected networks
of excitatory and inhibitory point neurons, in which firing activity
is described by the ``integrate-and-fire'' model, while synaptic
interactions are conductance-based.  As shown in previous
publications (Vogels and Abbott, 2005; El Boustani et al., 2007),
this model can display states of activity consistent with recordings
in awake neocortex, as shown in Fig.~\ref{network}.  The network can
display different types of states, such as ``synchronous regular''
(SR) states, or ``asynchronous irregular" (AI), both of which are
illustrated in Fig.~\ref{network}.  In SR states, the activity is
oscillatory and synchronized between neurons, while in AI states,
neurons are desynchronized and fire irregularly, similar to
recordings in awake cats (compare with cells in
Fig.~\ref{lfp-spikes}).  The averaged activity of the network is
coherent and of high amplitude in SR states, but is very noisy and of
low amplitude in AI states, similar to the EEG or LFP activity seen
in wakefulness (compare with EEGs in Fig.~\ref{eeg} and LFPs in
Fig.~\ref{lfp-spikes}).

\begin{figure}[h] 
\centerline{\psfig{figure=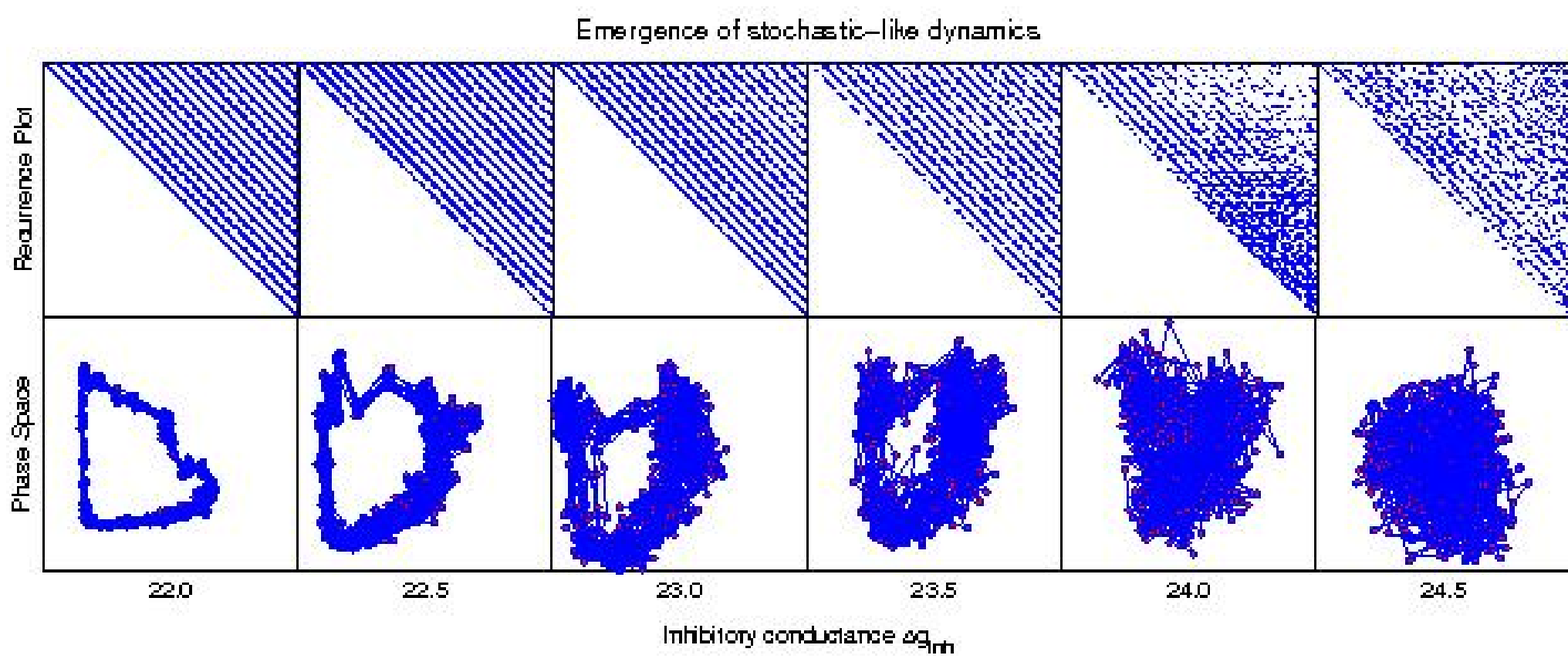,width=1.0\columnwidth}}

\caption{Recurrence plot and phase portraits of different network
states.} \label{recur}  From left to right, different step of the transition from the SR state (leftmost) to AI state (rightmost). All states differed by the value of the inhibitory conductance $\Delta g_{inh}$,
as indicated. The top graphs indicate the recurrence plot. The diagonals indicate the existence of a dominant periodic component. As the slaved degrees of freedom are unleashed with increasing inhibition, the recurrence plot is blurred by the stochastic-like component emerging at larger scales. This is the manifestation of the transition to a high-dimensional dynamics. For network states in the middle of the AI region, there is almost no visible recurrency and the recurrence plot looks like that of a stochastic process. The bottom graphs show bi-dimensional phase space reconstruction of the corresponding state. We can clearly see the "noisy" limit cycle which progressively degenerates into a dense high-dimensional attractor confirming what is shown by the recurrence plot.

\end{figure}

The time series of these averaged network activities (bottom graphs
in Fig.~\ref{network}) were analyzed similarly to the EEG in
Section~\ref{EEGsec}. We were particularly interested in the
transition region between AI and SR states where the firing looses
its coherence.  This trajectory in the phase diagram is shown as a
black rectangle in the upper panels of Fig.~\ref{network} and is
characterized by a abrupt drop of activity concomitant with an
increasing irregular firing. {Similar results can be obtained by
reducing the excitatory synaptic strength instead. However, to avoid
a loss of stability, we preferred to study the transition by
increasing the inhibitory synaptic strength.} Phase portraits of SR
and AI states, as well as the corresponding recurrence plots, are
illustrated in Fig.~\ref{recur}. As expected, SR states (leftmost
panel in Fig.~\ref{recur}) display a limit cycle phase portrait,
while AI states (rightmost panel) appear as a dense unstructured
attractor. When the dynamics is dominated by inhibition the limit
cycle is rapidly blurred by small-scale fluctuations which become
ubiquitous for AI states. However, even though the phase portrait
does not display any structure, the recurrence plot still exhibit a
strong deterministic component (diagonal line) which is completely
lost for a state well beyond the boundary region (noisy recurrence
plot).

\begin{figure}[h] 
\centerline{\psfig{figure=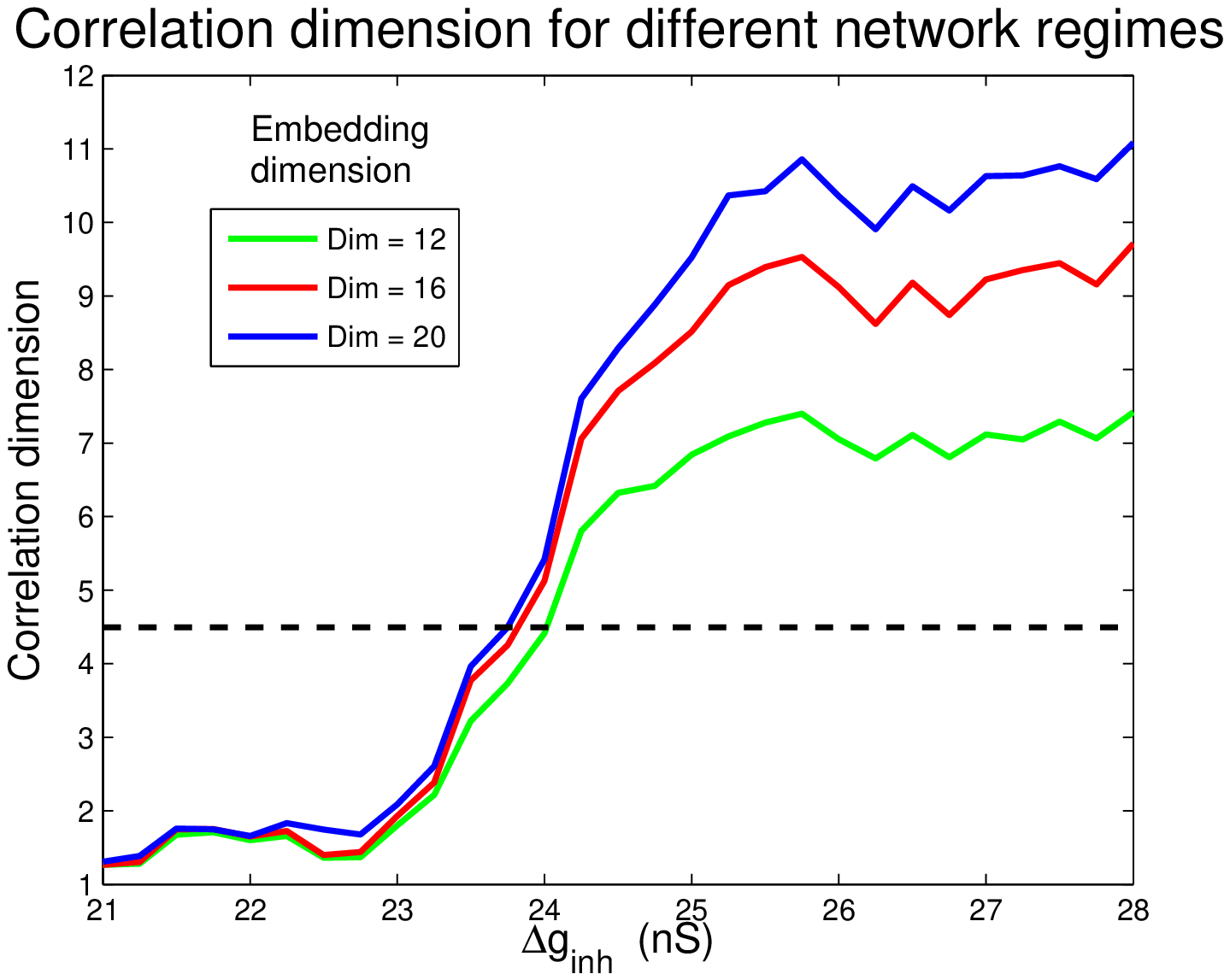,width=0.8\columnwidth}}

\caption{Transition to a high-dimensional attractor in the AI
regime.} \label{VAdim} The correlation dimension is plotted according
to the network state from the SR to the AI regime. Saturation of the
scaling region in the correlation integral is shown by illustrating
the estimated correlation dimension for three embedding dimensions.
Below the dotted line, the measure is considered to saturate
correctly whereas above this line, the measure can not be trusted
anymore. In particular for low inhibition conductances, there are
severe shifts with increasing embedding dimension. The dotted line is
given by the disappearance of the plateau in the $\epsilon$-entropy
of Fig.~\ref{lyapmodel}.

\end{figure}

To quantify this progressive release of degree of freedom, we estimated the correlation
dimension on the overall activity for those different states. Fig.~\ref{VAdim} shows the
computed dimension for three different embedding dimensions. The network activity displays a
low-dimension dynamics as expected for SR state whereas the dimension suddenly increases with
increasing inhibition conductance. Moreover, the corresponding values don't saturate with
increasing embedding dimension suggesting a high-dimensional attractor. Therefore, beyond 
the dotted black line, the correlation dimension is not a relevant measure anymore.

\begin{figure}[h] 
\centerline{\psfig{figure=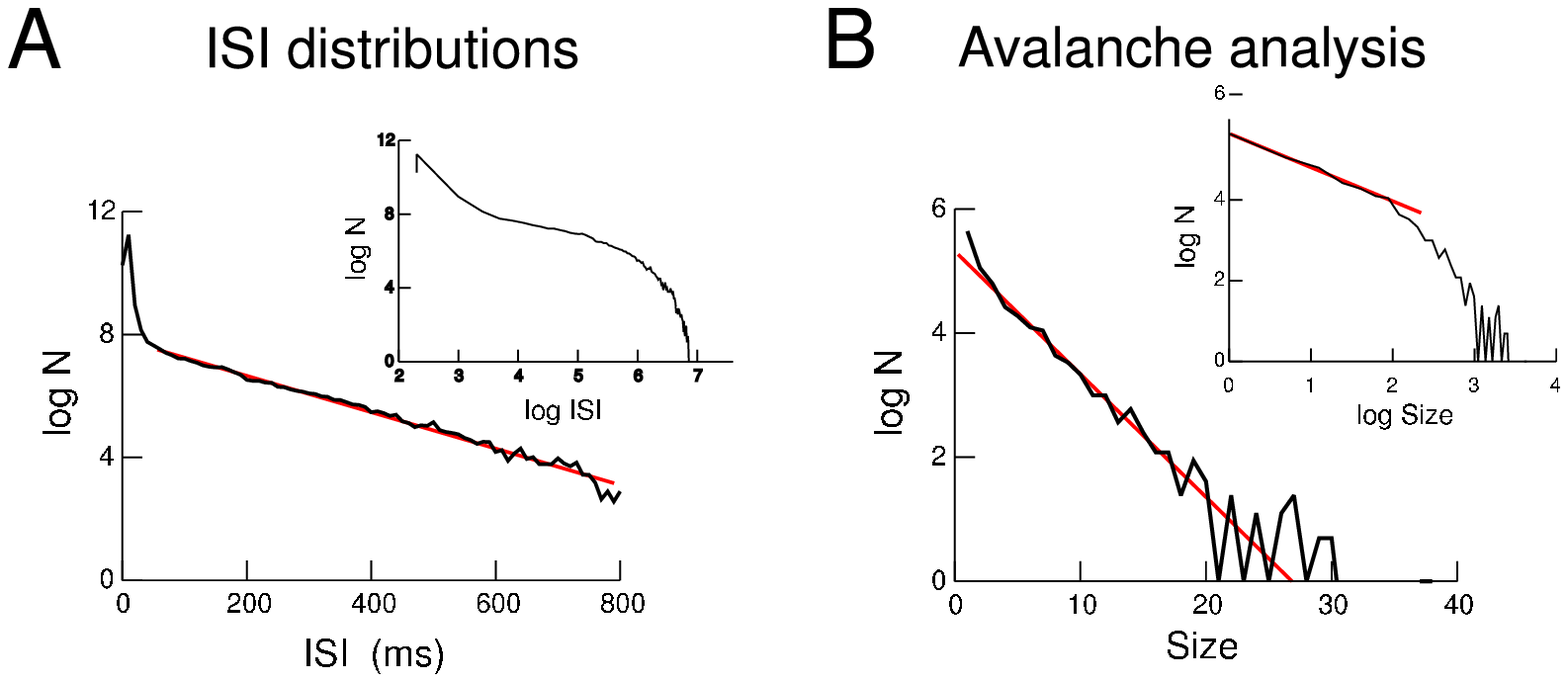,width=1\columnwidth}}

\caption{Analysis of the network dynamics in a model of asynchronous
irregular states.}  \label{AIanal}  A.  ISI distributions during the
AI state in a randomly-connected network of 16,000 integrate-and-fire
neurons.  ISI distributions are exponential (noisy trace), as
predicted by a Poisson process (red line; inset: log-log
representation using natural logarithms).  B.  Absence of avalanche
dynamics in this model (same description as Fig.~\ref{anal}B).  The
red lines indicates regions of exponential scaling; the red line in
inset indicates a region with power-law scaling. Modified from
El~Boustani et al., 2007.

\end{figure}

We also analyzed the spiking activity of the model using the same
statistical tools as for experimental data.  The distribution of ISIs
during AI states was largely dominated by an exponential component
(Fig.~\ref{AIanal}A, very similar to experimental data (compare with
Fig.~\ref{anal}A).  Analyzing population activity, through ``avalanche
analysis'', displayed exponential distributions (Fig.~\ref{AIanal}B),
also similar to experimental data (compare with Fig.~\ref{anal}B). 
Power-law scaling was also present for a limited range (see
Fig.~\ref{AIanal}B, inset).  Similar results were obtained for
different network states (not shown). 

{One can ask what would be the effect of a more specific
connectivity architecture on these results. In previous work, we have
shown that macroscopic properties in these networks are conserved for
more local connectivity, as long as the the connexions remain sparse
[El~Boustani \& Destexhe, 2009].  Therefore, the phase diagram for
topological networks owns similar structures as those displayed in
Fig.~\ref{network}. However, in the limit of ``first-neighbor''
connectivity, correlations between neurons become significantly
stronger and the resulting neuronal dynamics are more regular and far
from biological observations. Thus, even though more local
architectures can result in more correlated activity, these
correlations have to remain small in order to preserve the network
stability and the rich repertoire of regimes. Our results are thus
general for locally-connected networks as long as the sparseness is
respected.}

\section{How to reconcile these data ?}

The above data show that in both human, cat and models, the
dynamics can show clear signs of coherence and low dimensionality
at the level of macroscopic measurements (EEG, LFPs, averages),
while microscopically, neuronal dynamics are highly irregular and
resemble stochastic processes.  In an attempt to reconcile these
observations, we rely on recently introduced generalization of
classical nonlinear tools. In particular, Finite-Size Lyapunov
Exponent [Aurell et al., 1997; Shibata \& Kaneko, 1998; Cencini et
al., 1999; Gao et al., 2006] and the $\epsilon$-entropy [Cencini et
al., 2000] have proven to be valuable measures to probe system
which display different behaviors at different scales. In the
present context, where low-dimensional dynamics can take place in
top of a highly irregular neuronal behavior, those measures appear
as the most natural.  
 
\begin{figure}[h] 
\centerline{\psfig{figure=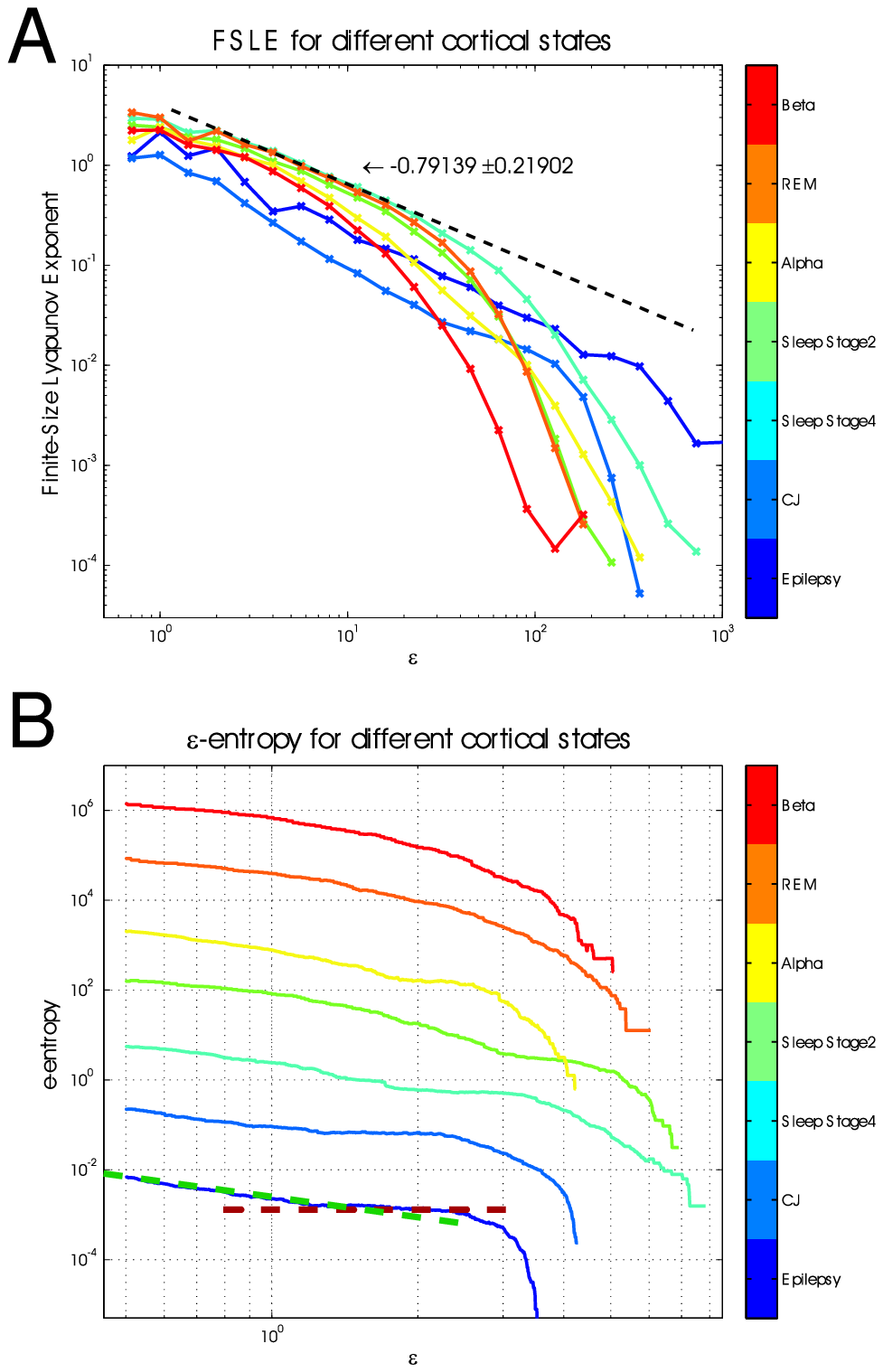,width=0.65\columnwidth}}

\caption{Scale-dependent Lyapunov exponent and Epsilon-entropy for
different brain states.} \label{lyapEEG}  The scale-dependent
Lyapunov exponent (A) and $\epsilon$-entropy (B) were calculated from
the same EEG states as shown in Fig.~\ref{eeg}.  The FSLE (A) does
not show any plateau and if the inverse FSLE is plotted in a semilog
scale on the x-axis, no low-dimensional chaotic behavior can be
detected for most states (data not shown). Following [Shibata \&
Kaneko,1998] example for general coupled maps, the macroscopic
activity seems to exhibit high-dimensional chaotic dynamics. For
small scales $\epsilon \rightarrow\ 0$, a scall-free region is found
with a power-law exponent around 0.791 (black dotted line) as
expected from the microscopic stochastic-like dynamics [Cencini et
al., 2000; Gao et al., 2006]. Because of the absence of a large
scale-free region in the semilog region, we can not define a
macroscopic Lyapunov exponent from this measure. However, for the
Creutzfeldt-Jacob disease and the epileptic seizure, the FSLE have a
free-scale region on a broad range which is different from the other
cortical states. This could indicate a low-dimensional chaotic
dynamic even though it is not reflected in the semilog scale. Indeed,
the $\epsilon$-entropy (B) manifests a clear plateau (red dotted
line) for the epileptic seizure, the Creutzfeldt-Jakob disease, sleep
stage 2 and 4 and the alpha waves. These plateau indicates the
existence of a low-dimensional attractor on the corresponding scales
[Cencini et al., 2000] which can sometimes be easily visualized with
an embedding procedure (see Fig.~\ref{eeg}). For REM and awake
states, this plateau disappears leaving the dynamics in a
stochastic-like (high-dimension) state at all scales. The small-scale
behavior is identical to the FSLE (green dotted line).

\end{figure}

They were applied to the human EEG, as shown in Fig.~\ref{lyapEEG}. 
The FSLE in the upper panel neither possess a plateau nor behave
according to a power-law except in the small-scale limit where the
stochastic-like component is dominant. For Creutzfeldt-Jacob disease
as well as for epileptic seizure, however, it seems that a small
region displays a almost scale-free behavior which would indicate a
low-dimensional chaotic dynamics. Following [Shibata \& Kaneko,
1998], it seems that the cortical activity lives in a
high-dimensional attractor which could be chaotic. Therefore, the
FSLE does not really help to untangle the different scales dynamics
here. However, when resorting to the $\epsilon$-entropy, we get a
different picture. Fig.~\ref{lyapEEG}B clearly shows that most of
these cortical states manifest a large plateau on large-scale.  These
plateau is a signature of low-dimensional dynamics [Cencini et al.,
2000] whereas the small-scale power-law behavior is the signature of
stochastic-like dynamics produced by high-dimensional attractor. In
accordance with Fig.~\ref{cdim}, {the dynamics during awake eyes
open and REM sleep do not own a plateau, and} their attractor
dimension is too high to be distinguished from noise. However, it
should be noted that those different dynamics are produced by the
same system hence being mainly modulated by endogenic factors. In
light of this result, we can conclude that there is no contradiction
between the experimental data acquired at the macroscopic level (EEG)
and the at the microscopic level (Spiking Activity). For large
network of coupled units, averaged quantities can display
low-dimensional structured dynamics while being seemingly
sto\-chastic at a smaller scale.

\begin{figure}[h] 
\centerline{\psfig{figure=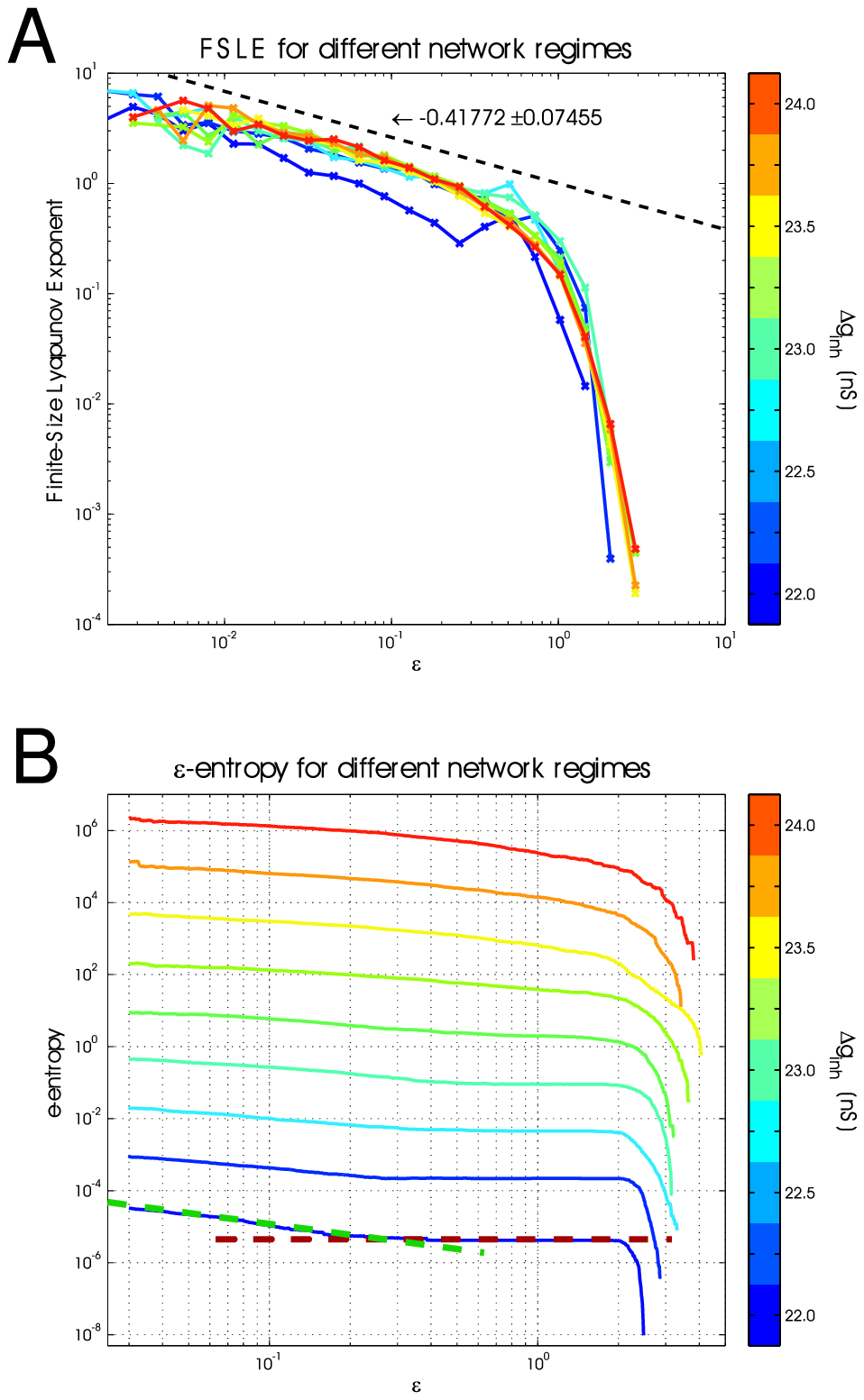,width=0.65\columnwidth}}

\caption{Scale-dependent Lyapunov exponent and Epsilon-entropy for
different network states.} \label{lyapmodel}  The Finite-Size
Lyapunov Exponent (A) and $\epsilon$-entropy (B) were calculated from
the same network states as shown in Fig.~\ref{network}. The FSLE
behaves almost identically for every network state. There is no
evidence for a low-dimensional chaotic dynamics which would be
indicated by a scale-free region in intermediate scales. The
power-law at small-scale owns a coefficient around 0.418 (black
dotted line) characterizing the stochastic-like dynamic
(high-dimensional deterministic) at those scales. Even though no
evidence for low-dimension chaos can be found from the FSLE, the
$\epsilon$-entropy (B) has a large plateau (red dotted line) for
state lying in the SR region (low inhibitory conductance). This
plateau is shortened when the network dynamics is driven toward the
AI regime where eventually the dynamics is indistinguishable from
stochastic process. The small-scale behavior is identical to the FSLE
(green dotted line).

\end{figure}

We also evaluated the same quantities from the averaged activities of
the numerical model. In Fig.~\ref{lyapmodel}A, we see as before that
the FSLE does not exhibit any scale-free region except for the
small-scale limit. Thus we can not rely on this measure to
distinguish small- and large- scale behavior. In contrast, the
$\epsilon$-entropy in Fig.~\ref{lyapmodel}B yields a large and
distinct plateau in the SR regime. The disappearance of this plateau
have been used as a criteria to draw the dotted black line in
Fig.~\ref{VAdim}. For inhibitory conductances larger than $\Delta
g_{inh} \simeq $ 23.5 (nS), the $\epsilon$-entropy slowly converges
to its small-scale stochastic-like behavior. We thus recover a very
comparable behaviour to the EEG data where pathological states or
deep sleep produce structured activity at the EEG level and awake
state or REM are comparable to the dynamics of AI regime, as
suggested previously [van Vreeswijk \& Sompolinsky, 2006; van
Vreeswijk \& Sompolinsky, 2008; Vogels \& Abbott, 2005; El Boustani
et al., 2007; Kumar et al., 2008].

\section{Discussion}

In this paper, we have briefly reviewed correlation dimension
analyses of human EEG, which revealed a hierarchy of brain states,
where the dimensionality varies approximately inversely to the level
of arousal (Fig.~\ref{dimampl}).  In particular, awake and attentive
subjects display high dimensionalities, while deep sleep of
pathological states show evidence for low dimensionalities.  These
low dimensions are difficult to reconcile with the fact that these
signals emanate from the activity of millions (if not billions) of
neurons.  In cat cerebral cortex, the cellular activity during wake
and sleep are highly irregular, and exponentially distributed like
stochastic (Poisson) processes, a feature which is also difficult to
reconcile with low dimensionalities at the EEG level, even though LFP
data do not manifest low-dimensional dynamics (data not shown).  We
performed similar analyses on computational models, which also
display these apparently coherent activities at the level of
large-scale averages, while the microscopic activity is highly
irregular.  In particular SR states can display coherent behavior at
large scales, while AI states do not show evidence for coherence,
similar to recordings in awake cats and humans.  

The nature of the dynamics exhibited by models is assimilable to
high-dimensional chaos.  It has been shown that AI states in such
models shut down after some time, and are thus transient in nature
[Vogels \& Abbott, 2005; El Boustani et al., 2007; Kumar et al.,
2008; El Boustani \& Destexhe, 2009].  This lifetime has been
estimated to increase exponentially with the network size [Kumar et
al., 2008; El Boustani \& Destexhe, 2009], and can reach considerable
times (beyond any reasonable simulation time; [Kumar et al., 2008]). 
Moreover, other recent studies [Cessac, 2008; Cessac \& Vi\'eville,
2008] have obtained analytical results on similar models where
``transient chaotic-like regimes'' were found.  More precisely, these
regimes are periodic, but with a period which also grows
exponentially with network size.  These results are reminiscent of
the non-attractive chaotic manifolds extensively discussed in the
literature [Crutchfield \& Kaneko, 1988; Dhamala et al., 2001;
Dhamala \& Lai, 2002; T\'el \& Lai, 2008].  Hence, even though the
chaotic nature of the dynamics is not inherent of the underlying
system, the network spends a long enough period trapped in this
transient dynamics indistinguishable from chaos from a numerical
point of view.

To characterize the dynamics at different scales, we estimated
quantities such as FSLE or $\epsilon$-entropy, which clearly show
that microscopic scales (neurons) tend to be very high-dimensional
and complex, in many ways similar to ``noise'', while more coherent
behavior can be present at large scales.  This analysis is consistent
with the recently proposed concept of ``macroscopic chaos'', where a
very high-dimensional microscopic dynamics coexists with low
dimensionality at the macroscopic level. In this paper, the chosen
numerical model is known to be purely deterministic and can display
highly irregular spiking patterns close to stochastic processes or
noise. From a nonlinear analysis point of view, most of this dynamics
is indistinguishable from a random process. However, as soon as a
large scale behaviour emerges, the $\epsilon$-entropy can keep track
of it and still manifests the small-scale irregular fluctuations.

We conclude that numerical models of recurrent neuronal networks,
with conductance-based integrate-and-fire neurons, can be assimilated
to high-dimensional chaotic systems, and are in many ways similar to
the EEG. Moreover, even though they exhibit limit-cycle regimes for
several parameter sets, this collective dynamics is built on top of
irregular and high-dimensional neuronal activity which is only
apparent at small-scales.  Interestingly, this scheme reminds fluid
dynamics, where a seemingly random microscopic dynamics may also
coexist with more coherent behavior at large scales.  This is the
case for example close to the transition to turbulence, where fluids
can show evidence for low-dimensional chaos [Brandstater et al.,
1983].  More developed turbulence, however, does not show such
evidence, presumably because a large number of degrees of freedom
have been excited and the high-dimensional dynamics is present at all
scales. {It is possible that similar considerations apply to
brain dynamics, especially when considering the recent debate about
the nature of the ongoing activity of visual primary cortex (V1). 
Recordings made using voltage-sensitive dyes imaging (VSDI) in this
cortical area have suggested that spontaneous activity consists of a
seemingly-random replay of sensory-evoked orientation maps of
activity in anesthetized [Kenet et al., 2003], but not in awake
animals [Omer et al., 2008].  These observations have raised the
question of whether the cortical activity of V1 could be described by
a single-state high-dimensional attractor or a low-dimensional
multistable attractor [Goldberg et al., 2004]. It is likely that at
the VSDI scale, both scenarios are possible at the same time but at
different scales, which would be consistent with the present results.}


\section*{Acknowledgments}

Research supported by the CNRS, ANR and the European Community
(FACETS grant FP6 15879).  More details are available at
\url{http://cns.iaf.cnrs-gif.fr}


\section*{References}

\begin{description}

\item Albers, D.J., Sprott, J.C. \& Crutchfield, J.P. [2006] "Persistent chaos in high dimensions," Phys. Rev. E \textbf{74}:057201

\item Aurell, E., Boffetta, G., Crisanti, A., Paladin, G. \& Vulpiani, A. [1997] "Predictability in the large : an extension of the concept of Lyapunov exponent," J. Phys. A \textbf{30}(1), 1-26

\item Babloyantz, A. \& Destexhe, A. [1986] "Low dimensional chaos in an instance of epileptic seizure", Proc. Natl. Acad. Sc. USA \textbf{83}, 3513-3517.


\item Babloyantz, A., Nicolis, C. \& Salazar, M. [1985] "Evidence for chaotic dynamics of brain activity during the sleep cycle", Phys. Lett. A \textbf{111}, 152-156.


\item {B\'edard, C., Kr\"oger, H. \& Destexhe, A. [2006] "Does
the 1/f frequency scaling of brain signals reflect self-organized
critical states?", Phys. Rev. Lett. \textbf{97}(11) : 118102}

\item Beggs, J. \& Plenz, D. [2003] Neuronal avalanches in neocortical circuits.  J. Neurosci. \textbf{23}, 11167-11177.

\item Bertschinger, N. \& Natschl\"ager, T. [2004] "Real-Time Computation at the Edge of Chaos in Recurrent Neural Networks," Neural Comp. \textbf{16}, 1413-1436

\item Brandstater, A., Swift, J., Swinney, H.L., Wolf, A., Farmer,
J.D., Jen, E. \& Crutchfield, J.P.  [1983] "Low-dimensional chaos in
a hydrodynamic system".  Phys. Rev. Lett. \textbf{51}: 1442-1445.

\item Cencini, M., Falcioni, M., Vergni, D. \& Vulpiani A. [1999] "Macroscopic chaos in globally coupled maps," Physica D \textbf{130}, 58-72

\item Cencini, M., Falcioni, M., Olbrich, E., Kantz, H. \& Vulpiani A. [2000] "Chaos or noise: Difficulties of a distinction," Physica D \textbf{130}, 58-72

\item Cessac, B. [2008] "A discrete time neural network model with spiking neurons; Rigorous results on the spontaneous dynamics," J. Math. Biol. \textbf{56}, 311-345

\item Cessac, B. \& Vi\'eville, T. [2008] "On Dynamics of Integrate-and-Fire Neural Networks with Conductance Based Synapses," Front. Comput. Neurosci. \textbf{2}(2)

\item Crutchfield, J.P. \& Kaneko, K. [1988] "Are attractors relevant to turbulence?," Phys. Rev. Lett. \textbf{60}(26), 2715-2718

\item Destexhe, A. [1992] {\it Nonlinear Dynamics of the Rhythmical Activity of the Brain} (in French), Doctoral Dissertation (Universit\'e Libre de Bruxelles, Brussels, 1992).  \url{http://cns.iaf.cnrs-gif.fr/alain\_thesis.html}

\item Destexhe, A., Contreras, D. \& Steriade, M. [1999] "Spatiotemporal analysis of local field potentials and unit discharges in cat cerebral cortex during natural wake and sleep states", J. Neurosci. \textbf{19}, 4595-4608.

\item Destexhe, A., Sepulchre, J.A. \& Babloyantz, A. [1988] "A
comparative study of the experimental quantification of deterministic
Chaos". Phys. Lett. A \textbf{132}: 101-106.

\item Dhamala, M., Lai, Y.-C. \& Holt, R.D. [2001] "How often are chaotic transients in spatially extended ecological systems?," Phys. Lett. A \textbf{280}, 297-302

\item Dhamala, M. \& Lai, Y.-C. [2002] "The natural measure of nonattracting chaotic sets and its representation by unstable periodic orbits," Int. J. Bif. Chaos \textbf{12}(12), 2991-3005

\item Eckmann, J.P., Kamphorst, S.O. \& Ruelle, D. [1987] "Recurrence Plots of Dynamical Systems," Europhys. Lett \textbf{5}, 973–977.

\item Elbert, T., Ray, W.J., Kowalik, Z.J., Skinner, J.E., Graf K.E. \& Birbaumer, N. [1994] "Chaos and physiology: deterministic chaos in excitable cell assemblies", Physiol. Rev. \textbf{74}: 1-47.

\item El Boustani, S., Pospischil, M., Rudolph-Lilith, M. \&
Destexhe, A. [2007] "Activated cortical states: experiments, analyses
and models", J. Physiol. Paris \textbf{101}: 99-109.

\item El Boustani, S. \& Destexhe, A. [2009] "A master equation
formalism for macroscopic modeling of asynchronous irregular activity
states," Neural Comp. \textbf{21}: 46-100

\item Frank, G.W., Lookman, T., Nerenberg, M.A.H., Essex, C. \&
Lemieux, J. [1990]  "Chaotic time series analysis of epileptic
seizures," Physica \textbf{46D}: 427-438.

\item Fraser, A.M. \& Swinney H.L. [1986] "Independent coordinates for strange attractors from mutual information," Phys. Rev. A \textbf{33}, 1134-1140

\item Gaspard, P. \& Wang, X.J. [1993] "Noise, chaos, and $(\tau, \epsilon)$-entropy per unit time," Phys. Rep. \textbf{235}(6), 321-373.

\item Goldberg, J.A., Rokni, U., \& Sompolinsky, H. [2004] "Patterns of ongoing activity and the functional architecture of the primary visual cortex," Neuron \textbf{42}, 489-500

\item Grassberger, P. \& Procaccia, I. [1983] "Measuring the
stran\-geness of strange attractors," Physica D \textbf{9}, 189–208

\item Gao, J.B., Hu, J., Tung, W.W. \& Cao Y.H. [2006] "Distinguishing chaos from noise by scale-dependent Lyapunov exponent," Phys. Rev. E \textbf{74}:066204

\item Hegger, R., Kantz, H. \& Schreiber, T. [1999] "Practical implementation of nonlinear time series methods: The TISEAN package," Chaos \textbf{9}(413)

\item Jensen, H.J. [1998] {\it Self-Organized Criticality.  Emergent Complex Behavior in Physical and Biological Systems.} (Cambridge University Press, Cambridge UK).


\item Kantz, H. \& Shreiber, T. [2004] "Nonlinear time series analysis," Cambridge University Press

\item Kenet, T., Bibitchkov, D., Tsodyks, M., Grinvald, A. \& Arieli, A. "Spontaneously emerging cortical representations of visual attributes," Nature \textbf{425}, 954-956

\item Korn, H. \& Faure, P. [2003] "Is there chaos in the brain? II. Experimental evidence and related models", C.R. Biol. \textbf{326}, 787-840.

\item Kumar, A., Schrader, S., Aertsen, A. \& Rotter, S. [2008] "The
High-Conductance State of Cortical Networks", Neural Comput.
\textbf{20}, 1-43



\item Legenstein, R. \& Maass, W. [2007] "Edge of chaos and prediction of computational performance for neural circuit models," Neural Net. \textbf{20}, 323-334

\item Marwan, N., Romano, M.C., Thiel, M. \& Kurths, J. [2006] "Recurrence plots for the analysis of complex systems" Physics Rep. \textbf{438}, 237:329

\item Mayer-Kress, G., Yates, F.E., Benton, L., Keidel, M., Tirsh,
W., Poppl, S.J. \& Geist, K. [1988]  "Dimension analysis of nonlinear
oscillations in brain, heart and muscle". Math. Biosci. \textbf{90}:
155-182.






\item Sauer, T., Yorke, J.A. \& Casdagli, M. [1991] "Embedology," J. Stat. Phys. \textbf{65}, 579-616


\item Shibata, T. \& Kaneko, K. [1998] "Collective chaos," Phys. Rev. Lett. \textbf{81}(9):4116

\item van Vreeswijk, C.A. \& Sompolinsky, H. [1996] "Chaos in neuronal networks with balanced excitatory and inhibitory activity," Science \textbf{274}, 1724-1726

\item van Vreeswijk, C.A. \& Sompolinsky, H. [1998] "Chaotic balanced state in a model of cortical circuits," Neural Comp. \textbf{10},1321-1372

\item Sompolinsky, H., Crisanti, A. \& Sommers,  H.J [1988] "Chaos in random neural networks,"  Phys. Rev. Lett. \textbf{61}, 259-262

\item Sprott, J.C. [2008] "Chaotic dynamics on large networks," Chaos \textbf{18}:023135 

\item Takens, F. [1981] "Detecting strange attractors in turbulence," Lecture notes in mathematics, Vol.898. Dynamical systems and turbulence, Springer, Berlin

\item Theiler, J. [1990] "Estimating fractal dimension," J. Opt. Soc. Amer. A \textbf{7}, 1055

\item T\'el, T. \& Lai, Y.-C. [2008] "Chaotic transients in spatially extended systems," Physics Rep. \textbf{460}, 245-275

\item Vogels, T.P. \& Abbott, L.F. [2005] "Signal propagation and logic gating in networks of integrate-and-fire neurons", J. Neurosci. \textbf{25}, 10786-10795.

\item Wolf, A., Swift, J.B., Swinney, H.L. \& Vastano, J.A. [1985] "Determining Lyapunov exponents from a time series", Physica D \textbf{16}, 285-317.

\end{description}

\end{document}